\documentclass{PoS}

\def\beq{\begin{equation}}
\def\eeq{\end{equation}}
\def\bea{\begin{eqnarray}}
\def\eea{\end{eqnarray}}

\newcommand{\uphi}{{\Phi}}
\newcommand{\ups}{{\Psi_1}}
\newcommand{\upS}{{\Psi_2}}

\title{Summary of Super Doubler Approach on Exact Lattice Supersymmetry}

\ShortTitle{Super Doubler Approach for Lattice SUSY}

\author{Alessandro D'Adda \\
        INFN Sezione di Torino, and\\
Dipartimento di Fisica Teorica,
Universita di Torino\\
I-10125 Torino, Italy\\
        E-mail: \email{dadda@to.infn.it}}

\author{\speaker{Noboru Kawamoto}\\
        Department of Physics, Hokkaido University\\
Sapporo, 060-0810 Japan\\
        E-mail: \email{kawamoto@particle.sci.hokudai.ac.jp}}

\author{Jun Saito \\
        Division of Natural Science
Department of Human Science\\
Obihiro University
of Agriculture and Veterinary Medicine\\
Obihiro, Japan\\
        E-mail: \email{jsaito@obihiro.ac.jp}}

\abstract{We have proposed a lattice SUSY formulation which we may 
call super doubler approach, where chiral fermion species doublers 
and their bosonic counter parts are either identified as super partners or 
truncated by chiral conditions. We claim 
that the super symmetry is exactly kept on the lattice. However 
the formulation is nonlocal and breaks lattice translational 
invariance. We argue that these features cause no fundamental 
difficulties in the continuum limit.  Although a naive version of this
formulation breaks associativity of the product of fields 
we have found a modified super doubler approach that recovers the 
associativity and is applicable to  super Yang-Mills theory. 
It turns out that this formulation is essentially equivalent to the 
continuum formulation and thus keeps all the symmetry exact  
even at a finite lattice constant. Inspired by this formulation we 
propose a non-local lattice field theory formulation which is free 
of chiral fermion problem and has the same exact lattice symmetry 
as continuum theory.}

\FullConference{The 33rd International Symposium on Lattice Field Theory\\
		14 -18 July 2015\\
		Kobe International Conference Center, Kobe, Japan*}

\begin{document}
\section{Introduction}
Although realization of exact SUSY on the lattice has a long history it is 
uncompleted subject\cite{Dondi-Nicolai}. We hope to find the clue 
for a new field theory formulation by understanding the fundamental 
difficulties to realize exact lattice SUSY. A part of the exact lattice 
SUSY of extended algebra was realized and extensively 
investigated\cite{Kaplan-,Damgaard-Matsu}. The link approach of 
super Yang-Mills theory was also proposed\cite{DKKN}. It turned 
out this formulation needs to be non-commutative and has Hopf 
algebraic symmetry\cite{DKS1}.

There are two major difficulties to realize exact supersymmetry on the 
lattice:\\
1) Lattice counter part of differential operator, a difference operator, 
in the SUSY algebra does not satisfy Leibniz rule which causes non-vanishing 
nature of surface terms. \\
2) Naive fermion formulation of chiral fermion generates fermion species 
doublers which breaks the balance of boson and fermion number. 
Any other fermion formulation may generate obstacles for realization of 
exact lattice SUSY due to the different treatment of fermions and bosons. 

Within the framework of locality, lattice translational invariance, and 
associativity, it was claimed that Leibniz rule of difference operator cannot 
be realized\cite{Kato-Sakamoto-So}. One may thus need to give up some of 
these principles to realize exact lattice SUSY. In fact it was pointed out 
that SLAC derivative would be the only solution compatible with the lattice 
SUSY version of Ginsparg-Wilson relation\cite{Bergner-Bruckman-}. It is well 
recognized that SLAC derivative is nonlocal derivative. It has also been 
pointed out that this type of nonlocality does not cause fundamental problem 
for the realization of SUSY on the lattice\cite{Bergner, Kado-Suzuki} 
Here we investigate a formulation to accept nonlocality but keep "exact SUSY" 
on the lattice.

In order to avoid chiral fermion species doublers we may introduce Wilson 
term and ask if one can obtain lattice SUSY invariance perturbatively.  
It was noticed that bosons also need corresponding terms to the Wilson terms 
of fermions\cite{Bartels-Krammer,Giedt-..}. Modern resolution of chiral 
fermion problem for QCD may give a suggestion that Ginsparg-Wilson 
fermion would solve lattice SUSY fermion treatment. It was pointed out 
that Majorana condition has a difficulty of compatibility with Ginsparg-WiIson 
relation\cite{Fujikawa}. After all in these formulations fermions are  
formulated differently from bosons the realization of exact lattice SUSY is 
difficult. 

Here we investigate a formulation of exact lattice SUSY invariant formulation 
within a framework of nonlocal field theory.  

\section{Modified lattice momentum conservation}

In order to realize lattice SUSY invariance it was suggested in the first 
pioneering paper by Dondi and Nicolai\cite{Dondi-Nicolai} that lattice 
momentum conservation could be modified as follows:
\begin{equation}
\delta(p_1+p_2+\cdots) \longrightarrow 
\delta\left( \frac{\sin ap_1}{a}+\frac{\sin ap_2}{a}+\cdots \right),
\end{equation}
where $\frac{\sin ap_\mu}{a}$ is the momentum representation of symmetric 
difference operator. We claim that this replacement is not enough to 
solve the second species doubler problem 2) mentioned the above. 

Let us consider what the meaning of this replacement is. We usually 
identify the lattice momentum  $-\frac{\pi}{a}\leq p_\mu\leq \frac{\pi}{a}$ 
as an angular variable identified from the momentum representation of 
difference operator $\frac{\sin ap_\mu}{a}$. This lattice momentum 
approaches to the continuum momentum $-\infty <\hat{p}_\mu<\infty$ 
in the continuum limit $a\rightarrow 0$. Since the lattice momentum 
is an angular variable it has different topological nature from the 
continuum momentum, which is the origin of chiral fermion species doubling. 
The most important advantage of this identification is that discrete lattice 
translation invariance is kept with this choice. We can, however, argue 
later that lattice translational invariance is not the mandatory requirement 
for the recovery of Poincare invariance in the continuum limit. 
Then we can equally well identify a lattice momentum $\Delta(p_\mu)$ as 
an corresponding lattice momentum if 
$\Delta(p_\mu)\rightarrow p_\mu ~(a\rightarrow 0)$.  

In contrast to the choice of $\Delta(p)=\frac{\sin ap}{a}$,  
what could be the best choice to solve the second problem 2) at 
the same time ? Coordinate representation of this symmetric 
difference operator which is Hermitian has the following well known 
form:  
\begin{equation}
i\Delta_s f(x) = i \frac{f(x+a)-f(x-a)}{2a} \rightarrow 
\frac{\sin ap}{a}\tilde{f}(p). 
\label{diff-op-norm}
\end{equation}
If we identify this as the translation generator the minimal unit of 
translation 
should be identified as two lattice unit. This point of view can be also 
understood by recognizing that species doubler state is a real physical 
state. We may define a new momentum for species doubler as 
$\frac{\sin a\left(\frac{\pi}{a}-p' \right)}{a} \sim p'~(a\rightarrow 0)$. The 
coordinate representation of the corresponding field is 
$\tilde{\psi}\left(\frac{\pi}{a}-p'\right) \rightarrow (-1)^{\frac{x}{a}} 
\psi(-x)$ 
where $x=na ~(n \in Z)$ is lattice coordinate. Translational invariant 
species doubler state is thus $(-1)^n const.$ which needs two lattice 
unit for translation invariance. 

The above observation suggests us a picture that we should construct 
a lattice formulation which leads single lattice as a minimal translation. 
In other words we need to introduce half lattice coordinates. Then 
what is the half lattice translation generator ? A half lattice translation 
can be nicely understood as supersymmetry translation, which is compatible 
with the following simplest SUSY algebra: $\{Q,Q\}=2P=2i\Delta_s$, 
where $i\Delta_s=\frac{2}{a}\sin \frac{pa}{2}$ is the momentum 
representation of single lattice translation generator while $Q$ is the 
half lattice translation generator. Our first step 
proposal for $\Delta(p)$ is 
\begin{equation}
 \Delta(p) = \frac{2}{a}\sin \frac{pa}{2}
 \label{lat-mom-1}
\end{equation}
We claim that this choice of $\Delta(p)$ solves two lattice SUSY 
difficulties 1) and 2) at the same time for non-gauge SUSY models 
where the following replacement need to be introduced:
\begin{equation}
\delta(p_1+p_2+\cdots ) \longrightarrow 
\delta(\Delta(p_1)+\Delta(p_2)+\cdots).
\label{lat-mom-con}
\end{equation}

In fact we have found momentum representation of lattice SUSY algebra 
for D=1,2 and N=2 Wess-Zumino models. In one dimension (D=1) we 
introduce bosonic and fermionic composite fields $\Phi(p)$ and $\Psi(p)$ 
each of which includes original field and species doubler fields for both 
bosonic and fermionic composite fields. SUSY transformation of the composite 
fields are 
\begin{eqnarray}
Q_1 \Phi(p)& =  i \cos \frac{a p}{4} \Psi(p),~~
Q_1 \Psi(p) = -4 i \sin \frac{a p}{4}  \Phi(p), \\ 
Q_2 \Phi(p)&=   \cos \frac{a p}{4}  \Psi(\frac{2\pi}{a}-p),~~
Q_2 \Psi(\frac{2\pi}{a}-p) = 4  \sin \frac{a p}{4}  \Phi(p),
\end{eqnarray}
which satisfy the following N=2 twisted SUSY algebra:
\begin{equation}
Q_1^2=Q_2^2=2\sin\frac{ap}{2}, ~~~~~~~~~~\{Q_1,Q_2\}=0,
\label{scharge-alg}
\end{equation}
where the dimensionless supercharges are introduced. 
We found the following action which is invariant under the above lattice SUSY 
algebra\cite{DFKKS}:
\begin{eqnarray}
 S^{(n)}&
 =g_0^{(n)}a^n \frac{4}{n!} \int_{-\frac{\pi}{a}}^{\frac{3\pi}{a}} \frac{d p_1}{2\pi} \cdots \frac{d p_n}{2\pi}
   2\pi\delta\left(\sum_{i=1}^n\sin\frac{a p_i}{2}\right)
  \nonumber\\
 &\qquad \times G(p_1,p_2,\cdots,p_n)\Bigl[
  2 \sin^2\frac{a p_1}{4} ~\Phi(p_1)\Phi(p_2)\cdots\Phi(p_n) + \label{kterm} \\
 & + \frac{n-1}{4}~ \sin\frac{a(p_1-p_2)}{4}~ \Psi(p_1)\Psi(p_2)\Phi(p_3)\cdots\Phi(p_n)\Bigr],
  \nonumber
\end{eqnarray}
where dimensionless $\Delta(p_i)=\sin \frac{ap_i}{2}$ is chosen as 
conserved lattice momentum of (\ref{lat-mom-con}) in this expression.
For $n=2$ this action gives kinetic and mass terms of one dimensional 
N=2 Wess-Zumino action. Here we identify the species doublers as 
super partners for boson and fermion:
\begin{eqnarray}
&& \Phi(p) = a^{-\frac{3}{2}} \varphi(p), ~~
 \Psi(p) = a^{-1} \psi_1(p), \nonumber\\
&& \Psi(\frac{2 \pi}{a} - p) = i a^{-1} \psi_2(p) , ~~
 \Phi(\frac{2 \pi}{a} - p) = -\frac{a^{-\frac{1}{2}}}{4} D(p).
\label{correspondence} 
\end{eqnarray} 

Similarly for D=N=2 lattice SUSY the kinetic terms of Wess-Zumino action 
can be found as:
\begin{eqnarray}
S_K &=& 4 \int_{-\frac{\pi}{a}}^{\frac{\pi}{a}} dp_+ dp_- dq_+ dq_-  
\delta(p_+ +q_+)\delta(p_- +q_-)
 \left[-4\bar{\uphi}(p) \sin\frac{aq_+}{2}\sin\frac{aq_-}{2}\uphi(q) 
\right.\nonumber \\
 &&\left. -\bar{F}(p) F(q)+2\bar{\upS}(p)\sin\frac{aq_+}{2} \upS(q) +
2\bar{\ups}(p)\sin\frac{aq_-}{2}\ups(q)\right],
\label{D-N-2-WZ-action2}
\end{eqnarray}
where for the kinetic term the proposed momentum conservation   
coincide with the standard lattice momentum conservation. 
The interaction terms for the Wess-Zumino action can be given by 
\begin{eqnarray}
S_n 
&=& \int \prod_{j=1}^n  d^2p_j V_n(p) n \left[i F(p_1)\prod_{j=2}^n \uphi(p_j) +(n-1) \upS(p_1)\ups(p_2) \prod_{j=3}^n \uphi(p_j) \right] +\textrm{h.c.},
\nonumber
\end{eqnarray}
where $V_n(p)$ includes sine momentum conservation:
\beq
V_n(p) = a^{2n}~ g_n~\delta^{(2)}\left(\sin\frac{ap_1}{2}+\sin\frac{ap_2}{2}+\cdots+\sin\frac{ap_n}{2}\right).
\label{sdelta}
\eeq
Chiral representation of D=N=2 SUSY algebra for the Wess-Zumino model 
can be fully derived with momentum representation of difference operator 
as momentum generator. See the details in \cite{DKKS}.


Chiral conditions impose the following relations for composite field of 
bosons and fermions $\uphi_A$, each of which include 4 species doublers:
\beq
\uphi_A(p_+,p_-) = \uphi_A(\frac{2\pi}{a}-p_+,p_-)=\uphi_A(p_+,\frac{2\pi}{a}-p_-)=\uphi_A(\frac{2\pi}{a}-p_+,\frac{2\pi}{a}-p_-).
\label{chiral2}
\eeq
For other composite fields $F(p), \Psi_i(p)$ similar chiral conditions 
are needed. The anti-chiral fields carrying bar need similar relations. 
In this way four species doubler degrees of freedom is all truncated by the 
chiral conditions. It is interesting to note that the coordinate representation 
of the first equality in eq.(\ref{chiral2}) is given as:
\beq
\Phi_A(x_+,x_-) = (-1)^{\frac{2x_+}{a}}\Phi_A(-x_+,x_-), 
\label{chiral-co}
\eeq
where the second and third equality can be given similarly. 
This relation shows that we have introduced twice larger degrees of 
freedom by introducing half lattice structure but it is truncated by half 
by the chiral condition for each dimension. It is thus sufficient to consider 
positive lattice coordinate for each dimensional direction.  

Since the species doubler degrees are killed now we can identify 
each composite fields as component fields: 
\beq
a^2 \uphi(p) \to \varphi(p),~~~~a^{\frac{3}{2}} {\Psi_i}(p) \to  \psi_i(p),
~~~~a F(p) \to  f(p).  \label{rescalingp}
\eeq
In this way for the treatment of species doubler degrees of freedom there are 
two possible ways:  A) Identify the species doubler fields as super partners 
as in one dimensional treatment in eq. (\ref{correspondence}). 
B) Kill the all the species doubler degrees of freedom by chiral conditions 
as in two dimensional treatment in eq.(\ref{chiral2}).
\section{Recovery of Leibniz rule and associativity}

We claim that the actions given above satisfy N=2 exact lattice SUSY 
algebra in one and two dimensions. How do we understand that the problem 
of difference operator not satisfying Leibniz rule can be realized in the 
coordinate space ?  Convolution of two product fields in the momentum 
space corresponds to the normal product in the coordinate space:
\begin{equation}
(F\cdot G)(p) = \int d^2 p_1 d^2 p_2 F(p_1) G(p_2)
\delta^{(2)}(p-p_1-p_2) ~ \longrightarrow ~~(F \cdot G)(x) = F(x) G(x). \label{dotprod}
\end{equation}
Correspondingly the change of the momentum conservation into sine 
of the momentum leads to define a new product: 
\begin{equation}
(F* G)(p) = \int d^2 p_1 d^2 p_2 F(p_1) G(p_2)
\delta^{(2)}(\Delta{(p)}-\Delta{(p_1)}-\Delta{(p_2)}) ~ \longrightarrow ~~(F * G)(x) = F(x)* G(x). \label{dotprod}
\end{equation}
If we now operate the momentum representation of difference operator 
to the product we naturally lead Leibniz rule in the momentum representation 
due to the new momentum conservation: 
\begin{equation}
 \Delta{(p)}~ (F*G)(p)
 = \int d^2 \hat{p}_1\, d^2 \hat{p}_2 \left[
\Delta{(p_1)} F(p_1) ~
G(p_2)+ F(p_1)\,\Delta{(p_2)}G(p_2) \right]
\delta^{(2)}(\Delta{(p)}-\Delta{(p_1)}-\Delta{(p_2)}.   \label{lbrule}
\end{equation}
The coordinate representation of this expression tells us that difference 
operator $\hat{\partial}_x$ satisfy the Libniz rule on the new $*-$product:
\begin{equation}
\hat{\partial}_x (F*G)(x) = \left( \hat{\partial}_x F(x) \right)* G(x) + F(x)   \left( \hat{\partial}_x *G(x) \right). \label{Lrule}
\end{equation}
Unfortunately this new $*-$product is unavoidably 
nonlocal\cite{Kato-Sakamoto-So}. The explicit expression of 
the $*-$product in the coordinate space can be found in 
\cite{DFKKS,DKKS}. 

It turns out, however, that this $*-$product is not associative due to the 
limited range of $|\Delta(p)|$. 
\begin{equation}
\left(({\Phi_3} *{\Phi_2})*{\Phi_1} \right)(p) \neq 
\left({\Phi_3} *({\Phi_2}* {\Phi_1}) \right)(p)
\label{asso}
\end{equation}
In order to formulate gauge theory associativity is crucial since gauge 
transformation is non-linear, which can be seen from the following 
manipulation proving gauge invariance:
\beq
\Phi^\dagger(x) * \Phi(x) \rightarrow (\Phi^\dagger(x)* e^{-\alpha(x)}) 
*(e^{\alpha(x)} \Phi(x)) \neq 
\Phi^\dagger(x)* (e^{-\alpha(x)}e^{\alpha(x)})\Phi(x) 
=  \Phi^\dagger(x) * \Phi(x).
\eeq

We have, however, found an interesting solution of the choice for 
$\Delta(p)=\Delta_G(p)$ which satisfies associativity:  
\begin{equation}
\Delta_G(p) = \frac{1}{a} \log 
\frac{1+ \sin \frac{ap}{2}}{1- \sin \frac{ap}{2}}, \label{gd}
\end{equation}
where $\Delta_G(\pm\frac{\pi}{a}) = \pm \infty$. The coordinate representation 
of this differential operator $\Delta_G$ has the following nice but 
non-local form:
\begin{equation}
\Delta_G\Phi(x) = \frac{2}{a} \sum_{k=1}^{\infty}  \frac{(-1)^{k+1}}{2k-1}  \left[ \Phi\left(x+\frac{(2k-1)a}{2}\right) -  \Phi\left(x-\frac{(2k-1)a}{2}\right)\right] \label{deltacoord}
\end{equation}

Replacing the $\Delta(p)$ in eq.(\ref{dotprod}) by $\Delta_G(p)$ we can 
define yet new $\star-$product which satisfies associativity:
\begin{equation}
\left(({\Phi_3} \star{\Phi_2}) \star{\Phi_1} \right)(p) =
\left({\Phi_3} \star({\Phi_2} \star{\Phi_1}) \right)(p).
\label{asso2}
\end{equation}
By the use of this formulation in principle we can formulate super 
Yang-Mills theory since lattice SUSY and gauge invariance could be assured. 

Since the lattice momentum conservation is changed as in 
eq.(\ref{lat-mom-con}) discrete lattice translational invariance is lost. 
However we claim that the invariance of the action for Poincare invariance 
can be assured by the corresponding invariance by 
\beq
\delta_\epsilon \Phi_A(p) = -i\epsilon \Delta_G(p) \Phi_A(p), 
\label{lat-trans}
\eeq
where newly defined $\Delta_G(p)$ can be identified as a translation 
generator. 
\section{Proposal of a new lattice field theory formulation} 
Based on the formulation of employing $\Delta_G(p)$ as a differential 
operator we propose to define a new lattice field theory formulation 
which is equivalent to continuum theory of paying  price of 
non-locality complication: \\
1) Introduce a half lattice structure. 
2) Go to the momentum representation of a continuum formulation and 
replace all derivative operators by $\Delta_G(p)$ and the momentum 
conservation in (\ref{lat-mom-con}) with $\Delta(p)=\Delta_G(p)$. 
3)  Kill the species doublers degrees of freedom by the equations as in 
(\ref{chiral2}). In the coordinate representation consider only the 
first quadrant of lattice space and replace all the product by 
$\star-$product. \\
In this way we can construct non-local lattice field theory which has 
exact lattice symmetries of continuum theory and no chiral fermion 
problem. In fact this non-local lattice theory is equivalent to the continuum 
theory. 

\section{Conclusion and Discussions}
We have proposed a lattice SUSY formulation which has the exact 
lattice SUSY especially for D=1,2 and N=2 Wess-Zumino models.
Even though associativity is broken in these cases exact lattice SUSY 
is kept since SUSY transformation is linear in fields. Non-associativity does 
not sacrifice exact lattice SUSY for non-gauge cases. In fact exact SUSY 
has been confirmed even at 
the quantum level\cite{Kondo-}. We have proposed a new lattice SUSY 
formulation by the use of $\Delta_G$ as a differential operator. It has the 
same exact lattice symmetries with continuum theory. In fact the formulation 
is non-local lattice field theory and equivalent to the continuum theory. 
In a sense this formulation presents "perfect action"\cite{Hasenfratz} 
of lattice SUSY. 
One may wonder if this lattice theory is regularized or not. We consider that 
the lattice SUSY formulated by $\Delta_G$ is not regularized 
even though it is a lattice theory with a lattice constant introduced. 
This is a puzzling situation.  If the non-local nature of the formulation is 
serious the formulation may be useless. We can,however, show that the 
locality of interactions is recovered in the continuum limit for both 
$\Delta(p)=\sin \frac{ap}{2}$ and $\Delta_G(p)$ formulations. 
The details will be given elsewhere\cite{DKS2}.

\noindent
{\bf Acknowledgements} \\
We thank P.H.Damgaard for useful discussions and comments. This work 
was supported in part by Japanese Ministry of Education, Science, Sports 
and Culture under the grant number 22540261 and also by the research 
funds of Instituto Nazionale di Fisica Nucleare (INFN).

\end{document}